\documentclass{article}
\pdfoutput=1 

\usepackage{natbib}
\usepackage{graphicx, setspace, amsmath, amssymb, subfigure, url, multirow, booktabs, verbatim, bm,threeparttable, amsthm, color}
\usepackage[linesnumbered,boxed,ruled,commentsnumbered]{algorithm2e}
\usepackage[toc]{appendix}
\usepackage{authblk}

\begin{document}




\title{Integrative Sparse Partial Least Squares}





\author{
  Weijuan Liang, Shuangge Ma, Qingzhao Zhang, Tingyu Zhu\thanks{Email: zhuti@oregonstate.edu}
}    

\date{03 July 2020}
\maketitle

\begin{abstract}
Partial least squares, as a dimension reduction method, has become increasingly important for its ability to deal with problems with a large number of variables.  Since noisy variables may weaken the performance of the model, the sparse partial least squares (SPLS) technique has been proposed to identify important variables and generate more interpretable results. However, the small sample size of a single dataset limits the performance of conventional methods. An effective solution comes from gathering information from multiple comparable studies. The integrative analysis holds an important status among multi-datasets analyses. The main idea is to improve estimation results by assembling raw datasets and analyzing them jointly. In this paper, we develop an integrative SPLS (iSPLS) method using penalization based on the SPLS technique. The proposed approach consists of two penalties. The first penalty conducts variable selection under the context of integrative analysis; The second penalty, a contrasted one, is imposed to encourage the similarity of estimates across datasets and generate more reasonable and accurate results. Computational algorithms are provided. Simulation experiments are conducted to compare iSPLS with alternative approaches. The practical utility of iSPLS is shown in the analysis of two TCGA gene expression data.
\end{abstract}

\section{Introduction}

With the rapid development of technology, comes the need to analyze data with high dimensions. Partial least squares, introduced by \citet{wold1984collinearity}, has been successfully used as a dimension reduction method in many research areas, such as chemometrics \citep{sjostrom1983multivariate} and more recently genetics \citep{Chun2009Expression}. PLS reduces the variable dimension by constructing new components, which are linear combinations of the original variables. Its stability under collinearity and high-dimensionality gives PLS a clear superiority over many other methods. However, in high dimensional problems, noise accumulation from irrelevant variables has long been recognized \citep{fan2010selective}. For example, in omics studies, it is wildly accepted that only a small fraction of genes are associated with outcomes. To yield more accurate estimates and facilitate interpretation, variable selection needs to be considered. Recently, \citet{chun2010sparse} propose a sparse PLS technique to conduct variable selection and dimension reduction simultaneously by imposing Elastic Net penalization in the PLS optimization. 

Another challenge that real data analyses often face is the unsatisfactory performance generated from a single dataset \citep{guerra2009meta}, especially for data with a limited sample size. Due to the recent progress in data collection, a possibility exists for integration across multiple datasets generated under similar protocols. Methods for analyzing multiple datasets include meta-analysis, integrative analysis, and others. 
Among them, integrative analysis has been proved to be effective both in theory and practice and have better performance in prediction and variable selection than other multi-datasets methods \citep{Liu2015Integrative,ma2011integrative}, especially including meta-analysis \citep{grutzmann2005meta}.

Considering the wide applications of PLS/SPLS to high dimensional data, we propose an integrative SPLS (iSPLS) method to remedy the aforementioned problems of the conventional SPLS technique caused by a limited sample size. Based on the SPLS technique, our method conducts the integrative analysis of multiple independent datasets using the penalization method to promote certain similarities and sparse structures among them, and further improve the accuracy and reliability of variable selection and loading estimation. Our penalization involves two parts. The first penalty conducts variable selection under the paradigm of integrative analysis \citep{Zhao2014Integrative}, where a composite penalty is adopted to identify important variables under both the homogeneity structure and heterogeneity structure. The intuition of adding the second penalty comes from empirical data analyses, that is, datasets with comparable designs may have a certain degree of similarity, which may help further improve analysis results. Our work advances from the existing sparse PLS and integrative studies by merging the dimension reduction technique and integrative analysis paradigm. Furthermore, we consider both similarity and difference across multiple datasets, which is achieved by our introduction of a two-part penalization. 

The rest of the paper is organized as follows. In Section 2, for the completeness of this article, we first briefly review the general principles of PLS and SPLS, and then formulate the iSPLS method and establish its algorithms. Simulation studies and applications to TCGA data are provided in Section 3 and 4. Discussion is organized in Section 5. Additional technical details and numerical results are provided in the Appendix.

\section{Methods}
\subsection{Sparse partial least squares}

Let $Y\in R^{n\times q} $ and $X\in R^{n\times p}$ represent the response matrix and predictor matrix, respectively. PLS assumes that there exists latent components $t_k$, $1\leq k \leq K$, which are linear combinations of predictors, such that  $Y=TQ^{\top}+F$ and $X=TP^{\top}+E$, where $T =(t_1,\dots, t_K)_{n\times K}$, $P\in R^{p\times K} $ and $Q\in R^{q\times K} $ are matrices of coefficients (loadings), and $E\in R^{n\times p} $ and $F\in R^{n\times q}$ are matrices of random errors. 

PLS solves the optimization problem for direction vectors $w_k$ successively. Specifically, $w_k$ is the solution to the following problem: 
\begin{equation} \label{eq:2.1}
\max \limits_{w_{k}}\{w_{k}^{\top}ZZ^{\top}w_{k}\}, 
 \end{equation}
which can be solved via the NIPALS \citep{wold1984collinearity} or SIMPLS \citep{de1993simpls} algorithms with different constraints and where $Z = X^{\top}Y$. After estimating the number of direction vectors $K$, the latent components can be calculated by $T=XW$,  where $W =(w_1,\dots, w_K)$.  And the final estimator is $\hat{\beta}^{PLS}= {W}_{K}\hat{Q}^{\top}$, where $\hat{Q}$ is the solution of $\min \limits_{Q}\{ \left\|Y-T_{K}Q^{\top}\right\|^{2}_{2}\}$. Details are available in \citet{ter1998objective}. 

In the analysis of high-dimensional data, a variable selection procedure needs to be considered to remove the noise. Note that noisy variables enter the PLS regression via direction vectors, one possible way is to adopt the penalization approach into the optimization procedure, that is, imposing an $\mathcal{L}_1$ constrain to the direction vector in problem (\ref{eq:2.1}). Then the first SPLS direction vector can be obtained by solving the following problem:
\begin{equation}\label{eq:2.2}
\max \limits_{w}\left\{w^{\top}ZZ^{\top}w\right\}, \quad \mbox{s.t.} ~w^{\top}w=1,\left| w \right| \leq \lambda, 
\end{equation} 
where the tuning parameter $\lambda$ controls the degree of sparsity.

However, \citet{jolliffe2003modified} point out the concavity issue of this problem as well as the lack of sparsity of its solution. \citet{chun2010sparse} then develop a generalized form of the SPLS problem (\ref{eq:2.2}) given below, which can generate a sufficiently sparse solution. 
 \begin{equation}\label{eq:2.3}
 \begin{aligned}
 \min \limits_{w, c} & \left\{-\kappa w^{\top}ZZ^{\top}w  +(1-\kappa)(c-w)^{\top}ZZ^{\top}(c-w) \right. 
 \left. +\lambda_{1}\left| c\right|_{1}+\lambda_{2}\left \| c \right \|_{2}^{2}\right\},\\
  &\mbox{s.t.} ~w^{\top}w=1.
 \end{aligned}
 \end{equation}
 In this problem,  penalties are imposed on $c$, a surrogate of the direction vector which is very close to $w$, rather than on the original direction vector. Here the additional $\mathcal{L}_{2}$ penalty deals with the singularity of $ZZ^{\top}$ when solving for $c$, and the small $\kappa$ reduces the effect of the concave part. The solution of (\ref{eq:2.3}) is given by optimizing $w$ and $c$ iteratively.

\subsection{Integrative sparse partial least squares}
\subsubsection{Data and Model Settings }

In this section, we consider the case where $L$ datasets are from independent studies with comparable designs. Below, we develop an integrative sparse partial least squares (iSPLS) method to conduct an integrative analysis of these $L$ datasets based on the SPLS technique. Note that in the context of integrative analysis, datasets do not need to be fully comparable. With matched predictors, we further assume that data preprocessing, including imputation, centralization, and normalization, has been done for each dataset separately. 

Following the notations in the existing integrative analysis literature \citep{huang2012identification,Zhao2014Integrative}, we use the superscript $(l)$ to denote the $l$th dataset $(Y^{(l)}_{n_l \times q}, X^{(l)}_{n_l \times p})$ with $n_l$ $i.i.d.$ observations, for $l=1, \dots, L$. As in the SPLS for a single dataset, where the main interest is on the first direction vector, denote $w_{j}^{(l)}$ as the weight of the $j$th variable in the first direction vector of the $l$th dataset, and $w_{j}=(w_{j}^{(1)}, \dots, w_{j}^{(L)})^{\top}$ as the “group” of weights of variable $j$ in the first $L$ direction vectors, for $j=1, \dots, p$

\subsubsection{iSPLS with contrasted penalization}
Following the generalized SPLS given in (\ref{eq:2.3}), we formulate the objective function for estimating the first direction vectors in $L$ datasets. For $l=1,\dots,L$, consider the minimization of the penalized objective function:
\begin{equation} \label{eq:2.4}  
\begin{aligned}
\sum_{l=1}^{L} &  \frac{1}{2n_l^2}\left(f(w^{(l)}, c^{(l)}) + \lambda\| c^{(l)} \|^{2}_{2} \right) +\text{pen}_{1}\left(c^{(1)}, \dots, c^{(L)}\right)
+\text{pen}_{2}\left(c^{(1)}, \dots, c^{(L)}\right)\\
& \mbox{s.t.} ~w^{(l)\top}w^{(l)}=1,
\end{aligned}
\end{equation}
where $f(w^{(l)}, c^{(l)})=-\kappa w^{(l)\top}Z^{(l)}Z^{(l)\top}w^{(l)}+(1-\kappa)(c^{(l)}-w^{(l)})^{\top}Z^{(l)}Z^{(l)\top}(c^{(l)}-w^{(l)})$, 
$c^{(l)} = (c_1^{(l)}, \cdots, c_{p}^{(l)})^\top$, $w^{(l)} = (w_1^{(l)}, \cdots, w_{p}^{(l)})^\top$ and $Z^{(l)}=X^{(l)\top}Y^{(l)}$ .

In (\ref{eq:2.4}),  $f(w^{(l)}, c^{(l)})$ is the goodness-of-fit of $l$th dataset, and $\| c^{(l)} \|^{2}_{2} $ serves the same role as in the SPLS method, dealing with the potential singularity when solving for $c^{(l)}$. To eliminate the influence of lager datasets, here we take the form of weighted sum with weights given by the reciprocal of the square of sample sizes. As for the penalty function, $\text{pen}_1(\cdot)$ conducts variable selection in the context of integrative analysis, whereas $\text{pen}_2(\cdot)$ accounts for the secondary model similarity structure. Below we provide detailed discussions on these two penalties.

\subsubsection{Penalization for variable selection }
We first consider the form of $\text{pen}_1(\cdot)$. With $L$ datasets, $L$ sparsity structures of the direction vectors need to be considered.  Integrative analysis considers two generic sparsity structures \citep{Zhao2014Integrative}, the homogeneity structure and the heterogeneity structure. Under the homogeneity structure, $I(w_{j}^{(1)}=0)=\dots =I(w_{j}^{(L)}=0)$, for any $j\in \{1, \dots, p\}$, which means that the $L$ datasets share the same set of important variables. Under the heterogeneity structure, for some $j \in \{1, \dots, p\}$, and $l,l^{\prime} \in \{1, \dots, L\}$, it is possible that $I(w_{j}^{(l)}=0)\not=I(w_{j}^{(l^{\prime})}=0)$, that is, one variable can be important in some datasets but irrelevant in others.

To achieve variable selection under the two sparsity structures, the composite penalty is used for $\text{pen}_1(\cdot)$, with the MCP as the outer penalty, which determines whether a variable is relevant at all. The minimax concave penalty (MCP)  is defined by $ \rho(t;\lambda,\gamma)=\lambda \int_{0}^{\left | t \right |}(1-x/(\lambda \gamma))_{+} \, dx$ \citep{Zhang2010Nearly} and its derivative $\dot \rho(t;\lambda,\gamma)=\lambda(1-\left |t \right|/(\lambda \gamma))_{+}\text{sgn}(t)$, where $\lambda$ is a penalty parameter, $\gamma$ is a regularization parameter that controls the concavity of $\rho$, $x_{+} = xI(x > 0)$, and $\text{sgn}(t)=-1,~ 0, ~\mbox{or}  ~1$ for $t< 0,~ = 0, ~\mbox{or} >0$, respectively. The inner penalties have different forms for the two sparsity structures. 

\paragraph{iSPLS under the homogeneity model}
Consider the penalty function 
 \begin{equation*} \text{pen}_{1}\left(c^{(1)} \dots, c^{(L)}\right)=\sum_{j=1}^{p}\rho\left(\left \|c_{j}\right \|_2;\mu_1, a\right),
 \end{equation*}
with regularization parameter $a$ and tuning parameter $\mu_1$. Here the inner penalty $\left\|c_{j}\right\|_2=\sqrt{\sum_{l=1}^{L}c_{j}^{(l)2}} $ is the $\mathcal{L}_{2}$ norm of $c_{j}$. Under this form of penalty, all the $L$ datasets select the same set of variables. The overall penalty is referred to as the 2-norm group MCP \citep{huang2012selective,ma2011integrative}.

\paragraph{iSPLS under the heterogeneity model}
Consider the penalty function 
 \begin{equation*}
 \text{pen}_{1}\left(c^{(1)} \dots, c^{(L)}\right)=\sum_{j=1}^p \rho \left( \sum_{l=1}^L \rho( | c_{j}^{(l)} |;\mu_1, a);1,b\right),
 \end{equation*}
with regularization parameters $a$ and $b$, and tuning parameter $\mu_1$. Here the inner penalty, which also takes the form of MCP, determines the individual importance for a selected variable. We refer to this penalty as the composite MCP.

\subsubsection{Contrasted penalization}
In the above section, the 2-norm MCP and composite MCP mainly conduct variable selection, but deeper relationships among datasets are ignored. It has been observed in empirical studies that, the estimation results of independent studies may exhibit a certain degree of similarity in their magnitudes or signs \citep{grutzmann2005meta,guerra2009meta}. It is quite possible that the direction vectors of the $L$ datasets have similarities in the magnitudes or signs if the datasets are generated by studies with similar designs \citep{guerra2009meta,shi2014integrative}. 

To utilize the similarity information and further improve estimation performance, we propose iSPLS with contrasted penalty $\text{pen}_2(\cdot)$, which penalizes the difference between estimators within each group. Specifically, we propose the following two kinds of contrasted penalties, depending on the degree of similarity across the datasets.

\paragraph{Magnitude-based contrasted penalization}
When datasets are quite comparable to each other, for example, those from the same study design but independently conducted, it is reasonable to expect that the first direction vectors have similar magnitudes. We propose a penalty which can shrink the differences of weights thus encourage the similarity within groups. Consider the magnitude-based contrasted penalty 
  \begin{equation*}
  \text{pen}_{2}\left(c^{(1)} \dots, c^{(L)}\right)= \frac{\mu_2}{2}\sum_{j=1}^p\sum_{ l^\prime \neq l}\left(c_{j}^{(l)}-c_{j}^{(l^\prime)}\right)^2,
  \end{equation*}
where $\mu_2 >0$ is a tuning parameter. Overall, we refer to this approach as iSPLS-Homo(Hetero)$_M$, with the subscript `M' standing for  magnitude. Here, we choose the $\mathcal{L}_2$ penalty for a simpler computation and note that it can be replaced by other penalties.

 \paragraph{Sign-based contrasted penalization}
Under certain scenarios, similarities in quantitative results is overly demanding, and it is more reasonable to expect/encourage the first direction vectors of the $L$ datasets to have similar signs \citep{Fang2018Integrative}, which is weaker than that in magnitudes. Here we propose the following sign-based contrasted penalty:
 \begin{equation*}
  \text{pen}_{2}\left(c^{(1)} \dots, c^{(L)}\right)= \frac{\mu_2}{2}\sum_{j=1}^p\sum_{l^\prime \neq l } \left\{\text{sgn}(c_{j}^{(l)})-\text{sgn}(c_{j}^{(l^\prime)})\right\}^2,
\end{equation*}
 where $\mu_2 >0$ is a tuning parameter, and $\text{sgn}(t)=-1, ~0$, or $1$ if $t< 0,~ = 0$, or $t >0$. Note that the sign-based penalty is not continuous, which brings challenges to optimization. We further propose the following approximation to tackle this non-smooth optimization problem:
  $$\frac{\mu_2}{2}\sum_{j=1}^p\sum_{ l^\prime \neq l } \Bigg ( \frac{c_{j}^{(l)}}{\sqrt{c_{j}^{(l)2}+\tau^2}}-\frac{c_{j}^{(l^\prime)}}{\sqrt{c_{j}^{(l^\prime)2}+\tau^2}} \Bigg )^2,$$
  where $\tau > 0$ is a small positive constant.

Under the `regression analysis + variable selection' framework, contrasted penalization methods similar to the proposed have  been developed \citep{Fang2018Integrative}. For the $j$th variable, the contrasted penalty encourages the direction vectors in different datasets to have similar magnitudes/signs, rather than forcing them to be the same. Even under the heterogeneity model, our two contrasted penalties are still reasonable. For example, they can encourage similarity within a group by pulling the nonzero loading which has relatively small value towards zero. The degree of similarity is adjusted by the tuning parameter $\mu_2$. Shrinkage of the differences between parameter estimates based on magnitude or sign has been considered in the literatures \citep{chiquet2011inferring,wang2016fused}, but is still novel under the context where we primarily focus on.

\subsection{Computation}
For the methods proposed in section 2.2, the computation algorithms share the same strategy with the SPLS procedure \citep{chun2010sparse}, where $w^{(l)}$ and $c^{(l)}$ are optimized iteratively for $l =1, \dots, L$. With fixed tuning and regularization parameters, the algorithm is repeated until convergence.  
\IncMargin{-0.3em} 
\begin{algorithm}
    \SetAlgoNoLine 
    Initialize. For $l = 1,\dots,L$:\linebreak 
      a. Apply partial least squares regression of $Y^{(l)}$ on $X^{(l)}$, and obtain the first direction vector $w^{(l)}$.\linebreak 
      b. Set $t=0$, $c_{[t]}^{(l)}=w_{[t]}^{(l)}=w^{(l)}$ and $Z^{(l)}=X^{(l)\top}Y^{(l)}$. \\
    Update:\linebreak 
      a. Optimize (\ref{eq:2.4}) over $w^{(l)}_{[t]}$ with fixed $c^{(l)}_{[t-1]}$.\linebreak 
      b. Optimize (\ref{eq:2.4}) over $c^{(l)}_{[t]}$ with fixed $w^{(l)}_{[t]}. $\\
    {Repeat Step 2 until convergence. In our simulation, we use the $\mathcal{L}_{2}$ norm of difference between two consecutive estimates smaller than a predetermined threshold as the criterion for convergence.} \\
     Normalize the final $c^{(l)}_{[t]}$ as $w^{(l)}=c^{(l)}_{[t]}/ \| c^{(l)}_{[t]} \|_2$ for each $l = 1,\dots,L$.\\
       \caption{Computational Algorithm for iSPLS}
\end{algorithm}
\DecMargin{-0.3em}

In Algorithm 1, the key  is Step 2. For Step 2(a), with fixed $c^{(l)}_{[t-1]}$, the objective function (\ref{eq:2.4}) becomes
\begin{equation}\label{eq:3.1}
\begin{aligned}
\min \limits_{w^{(l)}}\sum ^L_{l = 1}\left\{-\kappa w^{(l)\top}Z^{(l)}Z^{(l)\top}w^{(l)} \right. 
\left. +(1-\kappa)(c^{(l)}_{[t-1]}-w^{(l)})^{\top}Z^{(l)}Z^{(l)\top}(c^{(l)}_{[t-1]}-w^{(l)})\right\},
\end{aligned}
\end{equation}
which does not involve the group part. Thus, we can optimize $w^{(l)}$ in each dataset separately. Problem (\ref{eq:3.1}) can be written as 
\begin{equation*}
\begin{aligned}
\min\limits_{w^{(l)}} & \left\|Z^{(l)\top}w^{(l)}-\kappa^{\prime}Z^{(l)\top}c^{(l)}_{[t-1]}\right\|_{2}^{2},\quad \mbox{s.t.} ~w^{(l)\top}w^{(l)}=1,~\text{for} ~l=1,\dots,L, 
\end{aligned}
\end{equation*}
where $\kappa^{\prime}=(1-\kappa)/(1-2\kappa)$.  Then, by the method of Lagrangian multipliers, we have 
$$ w_{[t]}^{(l)}=\kappa^{\prime}(Z^{(l)}Z^{(l)\top}+\lambda^{*(l)}I)^{-1}Z^{(l)}Z^{(l)\top}c^{(l)}_{[t-1]},$$ 
where the multiplier $\lambda^{*(l)}$ is the solution of  $1/\kappa^{\prime2}=c^{(l)\top}_{[t-1]}Z^{(l)}Z^{(l)\top}(Z^{(l)}Z^{(l)\top}+\lambda I)^{-2}Z^{(l)}Z^{(l)\top}c^{(l)}_{[t-1]}$.\\

For Step 2(b), when solving $c^{(l)}$ for fixed $w_{[t]}^{(l)}$, problem (\ref{eq:2.4}) becomes
\begin{equation*}
\begin{aligned}
\min \limits_{c^{(l)}} \sum_{l=1}^L \frac{1}{2n_l^2} \left( \left\|Z^{(l)\top}c^{(l)}-Z^{(l)\top}w_{[t]}^{(l)}\right\|_{2}^{2}+\lambda \left \|c^{(l)}\right \|_{2}^{2} \right) &+\text{pen}_{1}\left(c^{(1)}, \dots, c^{(L)}\right)\\
&+\text{pen}_{2}\left(c^{(1)}, \dots, c^{(L)}\right).
\end{aligned}
\end{equation*}

The iSPLS algorithms under the homogeneity and heterogeneity models are different. We adopt the coordinate descent (CD) approach, which minimizes the objective function with respect to one group of coefficients at a time and cycles through all groups. This method transforms a complicated minimization problem into a series of simple ones. The remainder of this section describes the  CD algorithm for the heterogeneity model with a sign-based contrasted penalty. The computational algorithms for the homogeneity model and heterogeneity model with a magnitude-based contrasted penalty are described in the Appendix.

\subsubsection{iSPLS with the composite MCP }
 
Consider the heterogeneity model with the sign-based contrasted penalty, 
\begin{equation}
\begin{aligned}
\label{eq:3.2}
\min \limits_{c^{(l)}} &\sum_{l=1}^{L} \frac{1}{2n_l^2} \left( \left\|Z^{(l)\top}c^{(l)}-Z^{(l)\top}w_{[t]}^{(l)}\right\|_{2}^{2} 
+\lambda \left \|c^{(l)}\right \|_2^2\right) 
 + \sum_{j=1}^p \rho \left( \sum_{l=1}^L \rho(| c_{j}^{(l)} |;\mu_1, a); 1,b \right)\\
& + \frac{\mu_2}{2} \sum_{j=1}^p\sum_{l^\prime \neq l}\left\{\text{sgn}(c_{j}^{(l)})-\text{sgn}(c_{j}^{(l^\prime)})\right\}^2.
\end{aligned}
\end{equation}

For $j=1,\dots,,p$, given the group parameter vectors $c_{k}^{(l)}(k\not=j)$ fixed at their current estimates $c^{(l)}_{k,[t-1]}$, we minimize the objective function (\ref{eq:3.2}) with respect to $c^{(l)}_{j}$. $\lambda$ here is required to be very large because $Z^{(l)}$ is a $p \times q$ matrix with a relatively small $q$ \citep{chun2010sparse}. With $\lambda= \infty$, we take the first order Taylor expansion about $c_{j}^{(l)}$ for the first penalty, then the problem is approximately equivalent to minimizing 
\begin{equation*}
\begin{aligned}
 \frac{1}{2} & c^{(l)2}_{j} - w_{[t]}^{(l)\top}Z^{(l)}Z^{(l)\top}_{j}c^{(l)}_{j} + \alpha_{jl} | c^{(l)}_{j} | 
 + \frac{\mu_2^*}{2}\sum_{l^\prime \neq l } \Bigg ( \frac{c_{j}^{(l)}}{\sqrt{c_{j}^{(l)2}+\tau^2}}-\frac{c_{j,[t-1]}^{(l^\prime)}}{\sqrt{c_{j,[t-1]}^{(l^\prime)2}+\tau^2}} \Bigg )^2,
 \end{aligned}
\end{equation*}
where 
$\alpha _{jl}=\dot \rho (\sum_{l=1}^{L}\rho( | c_{j,[t-1]}^{(l)}|; \mu_{1}, a); 1, b )\dot\rho( | c^{(l)}_{j,[t-1]} |; \mu_{1}, a)$ and $\mu_2^*=\mu_{2}n_l^2$.

Thus, $c^{(l)}_{j,[t]}$ can be updated as follows: for $l=1,\dots,L$,\\
\begin{itemize}
\item[1.] Initialize $r=0$ and $  c^{(l)}_{j,[r]}= c^{(l)}_{j,[t-1]}$.
\item[2.] Update $r=r+1$. 
Compute:
 \begin{equation*}
  c^{(l)}_{j,[r]}=\frac{\text{sgn}(S_{j,[r-1]}^{(l)})(|S_{j,[r-1]}^{(l)}|-\alpha_{jl})_{+}}{(1+\mu_2^*(L-1))/( c^{(l)2}_{j,[r-1]}+\tau^{2})},
 \end{equation*}
  where
  \begin{equation*}
  \begin{aligned}
 S_{j,[r-1]}^{(l)}=&\sum_{m=1}^{p}\sum_{i=1}^{q}w^{(l)}_{m}Z^{(l)}_{mi}Z^{(l)}_{ji}
  +\frac{\mu_2^*}{\sqrt{c^{(l)2}_{j,[r-1]}+\tau^{2}}}\sum_{l^{\prime}\not=l}\frac{c_{j,[r-1]}^{(l^{\prime})}}{\sqrt{c_{j,[r-1]}^{(l^{\prime})2}+\tau^{2}}},
  \end{aligned}
  \end{equation*}
  and 
  $\alpha _{jl}=\dot \rho (\sum_{l=1}^L\rho( | c_{j,[r-1]}^{(l)}|; \mu_{1}, a); 1, b ) \dot\rho( | c^{(l)}_{j,[r-1]} |; \mu_{1}, a).$
\item[3.] Repeat Step 2 until convergence. The estimate at convergence is $c^{(l)}_{j,[t]}$.
\end{itemize}

\paragraph{Tuning parameter selection}
 iSPLS-Hetero$_S$ involves regularization parameters $a, b$. \citet{breheny2009penalized} suggested setting them connected in a manner to ensure that the group level penalty attains its maximum if and only if all of its components are at the maximum. Following published studies, we set $a = 6$. With the link between the inner and outer penalties, we set $b = \frac{1}{2}La\mu_1^2$. iSPLS-Homo$_S$ only involves regularization parameters $a$, which is also set to be 6. We use cross-validation to choose tuning parameters $\mu_1$ and $\mu_2$. Furthermore, iSPLS-Hetero$_S$ involves $\tau$. In our study, we fix the value of $\tau^2=0.5$, following the suggestion of setting it as a small positive number \citep{dicker2013variable}. Literature suggested that the proposed approach is valid if $\tau$ is not too big, and the approximation can differentiate parameters with different signs.

\section{Simulation}

We simulate four independent studies each with sample size 40 and 120, and 5 response variables.  For each sample, we simulate 100 predictor variables, which are jointly normally distributed, with marginal means zero and variances one. We assume that the predictor variables have an auto-regressive correlation structure, where variables $j$ and $k$ have correlation coefficient $\rho^{\left | j-k\right |}$, and $\rho=$ 0.2 and 0.7, corresponding to weak and strong correlations, respectively. All the scenarios follow the model $Y^{(l)} =X^{(l)}\beta^{(l)} +\epsilon^{(l)}$, where $\epsilon^{(l)}$ is normally distributed with mean zero. Following the data-generating mechanism in \citet{chun2010sparse}, the columns of $\beta_i^{(l)}$, for $i=2,...,5$, are generated by $\beta_i^{(l)}=1.2^{i-1}{\beta}_1^{(l)}$. The sparsity structures of direction vectors $w^{(l)}$ are controlled by $\beta_1^{(l)}$. Within each dataset, the number of variables associated with the responses is set to be 10. The nonzero coefficients $\beta_1^{(l)}$ range from 0.5 to 4. We simulate under both the homogeneity and heterogeneity models. 
 
Under the homogeneity model, direction vectors have the same sparsity structure, with similar or different nonzero values, corresponding to Scenario 1 and Scenario 2, respectively. Under the heterogeneity model,  two scenarios are considered. In Scenario 3, four datasets share 5 important variables in common, and the rest important variables are dataset-specific. That is, direction vectors have partially overlapping sparsity structures. In Scenario 4, direction vectors have random sparsity structures with random overlappings. These four scenarios comprehensively cover different degrees of overlapping in sparsity structures.
  
To better gauge performance of the proposed approach, we also consider the following alternative approaches: (a) meta-analysis. We analyze each data set separately using the PLS or SPLS approaches and then combine results across datasets via meta-analysis; (b) a pooled approach. Four datasets are pooled together and analyzed by SPLS as a whole. For all approaches, the tuning parameters are selected via 5-fold cross-validation. To evaluate the accuracy of variable selection, the averages of sensitivities and specificities are computed across replicates. We also evaluate prediction performance by calculating mean-squared prediction errors (MSPE).
    
Summary statistics based on 50 replicates are presented in Tables 1-4. The simulation indicates that the proposed integrative analysis method outperforms its competitors. More specifically, under the fully overlapping (homogeneity) case, when the magnitudes of nonzero values are similar across datasets (Scenario 1), iSPLS-Homo$_M$ has the most competitive performance. For example, in Table 1, with $\rho=0.2$ and $n=120$, MSPEs are 49.062 (meta-PLS), 5.686 (meta-SPLS), 1.350 (pooled-SPLS), 2.002 (iSPLS-Homo$_M$), 2.414 (iSPLS-Homo$_S$), 3.368 (iSPLS-Hetero$_M$) and 3.559 (iSPLS-Hetero$_S$), respectively. Note that under Scenario 1, the performance of iSPLS-Homo$_M$ and iSPLS-Homo$_S$ may be slightly inferior to that of pooled-SPLS. Since with fully comparable datasets, it is sensible to pool all data together, thus, pooled-SPLS may generate more accurate results. However, when the nonzero values are quite different across datasets (Scenario 2), as can be seen from Table 2, iSPLS-Homo$_S$ outperforms others, including pooled-SPLS. Under the partially overlapping Scenario 3 (heterogeneity model), iSPLS-Hetero$_M$ and  iSPLS-Hetero$_S$ seem to have better performance, for example when $\rho=0.7$ and $n=40$, they have higher Sensitivities (0.821 and 0.821, compared to  0.675, 0.575, 0.800 and 0.800 of the alternatives), smaller MSPEs (24.637 and 23.734, compared to  268.880, 30.928, 84.875, 40.867, and 39.492 of the alternatives), and with similar Specificities. Even under the non-overlapping Scenario 4, which is not favourable to multi-datasets analysis, the proposed integrative analysis still has reasonable performance. Thus, our integrative analysis methods have the potential to generate more satisfactory results comparable to meta-analysis, when the overlapping structure of multiple datasets is unknown.

\begin{table}
	\caption{Simulation results for Scenario 1($M=4, p=100$)}
	\begin{center}
	\label{Table1}
	\resizebox{125mm}{50mm}{
	\centering{\scriptsize
		\begin{tabular}{llllllllllll}
		\hline\

		        &$\rho$&$n_l$&Method&MSPE&Sensitivity&Specificity& \\
		       \hline
	
		       &0.2&40& meta-PLS & 48.972 (4.676) & 1 (0) & 0 (0)&\\
		                  &&& meta-SPLS & 24.739 (3.879) & 0.632 (0.135) & 0.873 (0.111)&\\
		                  &&& pooled-SPLS & 4.377 (2.486) & 0.810 (0.127) & 0.999 (0.003)&\\
		                  &&& iSPLS-Homo$_M$& 9.452 (4.369) & 0.840 (0.110) & 0.982 (0.018)&\\
		                  &&& iSPLS-Homo$_S$& 10.151 (4.027) & 0.837 (0.119) & 0.980 (0.022)&\\
		                  &&& iSPLS-Hetero$_M$& 18.287 (6.022) & 0.845 (0.152) & 0.757 (0.063)&\\
		                  &&& iSPLS-Hetero$_S$& 15.462 (6.251) & 0.875 (0.143) & 0.743 (0.060)&\\

		                 \hline
                        & 0.2&120& meta-PLS & 49.062 (4.151) & 1 (0) & 0 (0)&\\
		                  &&& meta-SPLS & 5.686 (2.056) & 0.799 (0.053) & 0.994 (0.007)&\\
		                  &&& pooled-SPLS & 1.350 (1.229) & 0.937 (0.025) & 0.999 (0.000)&\\
		                  &&& iSPLS-Homo$_M$ & 2.002 (0.920) & 0.993 (0.008) & 0.956 (0.016)&\\
		                  &&& iSPLS-Homo$_S$& 2.414 (0.951) & 0.997 (0.008) & 0.929 (0.014)&\\
		                  &&& iSPLS-Hetero$_M$& 3.368 (1.211) & 0.955 (0.039) & 0.945 (0.019)&\\
		                  &&& iSPLS-Hetero$_S$ & 3.559 (1.297) & 0.982 (0.051) & 0.872 (0.007)&\\
		          \hline
		       &0.7&40& meta-PLS & 106.532 (7.066) & 1 (0) & 0 (0)&\\
		                  &&& meta-SPLS & 16.212 (4.033) & 0.828 (0.063) & 0.962 (0.011)&\\
		                  &&& pooled-SPLS & 5.984 (1.939) & 0.893 (0.065) & 0.984 (0.037)&\\
		                  &&& iSPLS-Homo$_M$& 6.956 (1.885) & 0.967 (0.018) & 0.947 (0.021)&\\
		                  &&& iSPLS-Homo$_S$& 7.000 (2.067) & 0.967 (0.018) & 0.946 (0.020)&\\
		                  &&& iSPLS-Hetero$_M$& 13.630 (3.817) & 0.896 (0.109) & 0.946 (0.019)&\\
		                  &&& iSPLS-Hetero$_S$ & 13.855 (3.778) & 0.909 (0.112) & 0.942 (0.020)&\\
		                 \hline
		      &0.7&120& meta-PLS & 102.629 (9.225) & 1 (0) & 0 (0)&\\
		                  &&& meta-SPLS & 4.824 (1.913) & 0.912 (0.049) & 0.985 (0.012)&\\
		                  &&& pooled-SPLS & 2.454 (1.481) & 0.883 (0.056) & 0.994 (0.023)&\\
		                  &&& iSPLS-Homo$_M$& 2.292 (0.829) & 0.987 (0.018) & 0.977 (0.014)&\\
		                  &&& iSPLS-Homo$_S$& 2.356 (0.785) & 0.987 (0.018) & 0.976 (0.014)&\\
		                  &&& iSPLS-Hetero$_M$& 3.718 (0.995) & 0.988 (0.051) & 0.948 (0.013)&\\
		                  &&& iSPLS-Hetero$_S$& 3.609 (1.077) & 0.997 (0.035) & 0.942 (0.012)&\\
		                  	          
		\hline
			
		\end{tabular}
	}
	}
\end{center}
\end{table}

\begin{table}
	\caption{Simulation results for Scenario 2($M=4, p=100$)}
	\begin{center}
	\label{Table2}
	\resizebox{125mm}{50mm}{
	\centering{\scriptsize
		\begin{tabular}{llllllllllll}
		\hline\

&$\rho$&$n_l$&Method&MSPE&Sensitivity&Specificity& \\
		       \hline
		
		       &0.2&40& meta-PLS & 87.769 (14.532) & 1 (0) & 0 (0)&\\
		                  &&& meta-SPLS & 31.173 (8.422) & 0.532 (0.074) & 0.919 (0.073)&\\
		                  &&& pooled-SPLS & 33.533 (4.519) & 0.883 (0.095) & 0.976 (0.042)&\\
		                  &&& iSPLS-Homo$_M$& 17.567 (5.086) & 0.993 (0.025) & 0.681 (0.084)&\\
		                  &&& iSPLS-Homo$_S$ & 16.881 (4.548) & 0.993 (0.025) & 0.681 (0.084)&\\
		                  &&& iSPLS-Hetero$_M$& 28.803 (6.574) & 0.756 (0.122) & 0.774 (0.057)&\\
		                  &&& iSPLS-Hetero$_S$ & 25.990 (5.446) & 0.819 (0.102) & 0.739 (0.063)&\\
		                 \hline
                        & 0.2&120& meta-PLS & 85.138 (4.172) & 1 (0) & 0 (0)&\\
		                  &&& meta-SPLS & 9.015 (1.283) & 0.672 (0.054) & 0.994 (0.005)&\\
		                  &&& pooled-SPLS & 27.068 (0.867) & 0.993 (0.076) & 1.000 (0.003)&\\
		                  &&& iSPLS-Homo$_M$& 3.673 (0.552) & 1.000 (0.025) & 0.983 (0.024)&\\
		                  &&& iSPLS-Homo$_S$ & 3.589 (0.649) & 1.000 (0.018) & 0.982 (0.043)&\\
		                  &&& iSPLS-Hetero$_M$ & 6.050 (0.555) & 0.898 (0.040) & 0.956 (0.024)&\\
		                  &&& iSPLS-Hetero$_S$ & 6.674 (0.776) & 0.949 (0.030) & 0.939 (0.032)&\\		          
		                  \hline
		       &0.7&40& meta-PLS & 192.366 (10.990) & 1 (0) & 0 (0)&\\
		                  &&& meta-SPLS & 28.179 (5.592) & 0.652 (0.078) & 0.981 (0.015)&\\
		                  &&& pooled-SPLS & 65.284 (5.221) & 0.970 (0.101) & 0.963 (0.018)&\\
		                  &&& iSPLS-Homo$_M$& 10.186 (4.096) & 0.997 (0.055) & 0.948 (0.023)&\\
		                  &&& iSPLS-Homo$_S$& 9.909 (4.031) & 0.997 (0.055) & 0.947 (0.022)&\\
		                  &&& iSPLS-Hetero$_M$& 23.300 (9.108) & 0.741 (0.096) & 0.953 (0.017)&\\
		                  &&& iSPLS-Hetero$_S$& 22.974 (9.806) & 0.765 (0.095) & 0.950 (0.019)&\\		                          
		                  \hline
		      &0.7&120& meta-PLS & 175.348 (8.390) & 1 (0) & 0 (0)&\\
		                  &&& meta-SPLS & 14.871 (1.943) & 0.745 (0.041) & 0.975 (0.010)&\\
		                  &&& pooled-SPLS & 61.626 (0.758) & 0.963 (0.059) & 0.986 (0.008)&\\
		                  &&& iSPLS-Homo$_M$& 5.923 (0.913) & 0.997 (0.035) & 0.972 (0.012)&\\
		                  &&& iSPLS-Homo$_S$ & 5.764 (0.913) & 0.997 (0.035) & 0.971 (0.013)&\\
		                  &&& iSPLS-Hetero$_M$ & 10.742 (1.252) & 0.911 (0.016) & 0.917 (0.012)&\\
		                  &&& iSPLS-Hetero$_S$& 9.354 (1.267) & 0.946 (0.009) & 0.912 (0.011)&\\
		                  	          
		\hline
			
		\end{tabular}
	}
	}
\end{center}
\end{table}

\begin{table}
	\caption{Simulation results for Scenario 3($M=4, p=100$)}
	\begin{center}
	\label{Table3}
	\resizebox{125mm}{50mm}{
	\centering{\scriptsize
		\begin{tabular}{llllllllllll}
		\hline\

&$\rho$&$n_l$&Method&MSPE&Sensitivity&Specificity& \\
		       \hline

		       &0.2&40& meta-PLS & 76.919 (11.918) & 1 (0) & 0 (0)& \\
		                  &&& meta-SPLS & 35.372 (6.920) & 0.551 (0.118) & 0.900 (0.087)& \\
		                  &&& pooled-SPLS & 54.006 (6.920) & 0.675 (0.289) & 0.730 (0.289)& \\
		                  &&& iSPLS-Homo$_M$ & 28.495 (4.416) & 0.900 (0.057) & 0.589 (0.062)& \\
		                  &&& iSPLS-Homo$_S$ & 27.897 (4.231) & 0.900 (0.069) & 0.589 (0.070)& \\
		                  &&& iSPLS-Hetero$_M$ & 23.201 (5.788) & 0.800(0.137) & 0.847 (0.042)& \\
		                  &&& iSPLS-Hetero$_S$ & 21.616 (5.632) & 0.800(0.134) & 0.856 (0.039)& \\
		                 \hline
                        & 0.2&120& meta-PLS & 84.613 (16.931) & 1 (0) & 0 (0)& \\
		                  &&& meta-SPLS & 10.995 (2.382) & 0.696 (0.082) & 0.990 (0.010)& \\
		                  &&& pooled-SPLS & 44.243 (3.532) & 0.600 (0.190) & 0.847 (0.125)& \\
		                  &&& iSPLS-Homo$_M$ & 12.445 (1.995) & 0.902 (0.050) & 0.683 (0.049)& \\
		                  &&& iSPLS-Homo$_S$ & 12.471 (1.993) & 0.908 (0.049) & 0.674 (0.058)& \\
		                  &&& iSPLS-Hetero$_M$ & 8.699 (1.768) & 0.882 (0.050) & 0.926 (0.016)& \\
		                  &&& iSPLS-Hetero$_S$ & 8.467 (1.826) & 0.882 (0.049) & 0.931 (0.015)& \\
	                    \hline
		       &0.7&40& meta-PLS & 268.880 (12.323) & 1 (0) & 0 (0)& \\
		                  &&& meta-SPLS & 30.928 (7.532) & 0.675 (0.084) & 0.939 (0.022)& \\
		                  &&& pooled-SPLS & 84.875 (7.594) & 0.575 (0.152) & 0.909 (0.073)& \\
		                  &&& iSPLS-Homo$_M$ & 40.867 (6.147) & 0.800 (0.084) & 0.700 (0.152)& \\
		                  &&& iSPLS-Homo$_S$ & 39.492 (5.919) & 0.800 (0.080) & 0.700 (0.171)& \\
		                  &&& iSPLS-Hetero$_M$ & 24.637 (6.887) & 0.821 (0.102) & 0.900 (0.051)& \\
		                  &&& iSPLS-Hetero$_S$& 23.734 (6.373) & 0.825 (0.111) & 0.911 (0.068)& \\
		                  \hline
		      &0.7&120& meta-PLS & 258.583 (8.390) & 1 (0) & 0 (0)& \\
		                  &&& meta-SPLS & 12.631 (2.791) & 0.900 (0.062) & 0.971 (0.011)& \\
		                  &&& pooled-SPLS & 73.999 (6.112) & 0.800 (0.138) & 0.772 (0.135)& \\
		                  &&& iSPLS-Homo$_M$ & 20.475 (3.493) & 0.998 (0.010) & 0.364 (0.066)& \\
		                  &&& iSPLS-Homo$_S$ & 20.463 (3.445) & 0.998 (0.010) & 0.364 (0.066)& \\
		                  &&& iSPLS-Hetero$_M$ & 10.228 (2.837) & 0.988 (0.019) & 0.895 (0.022)& \\
		                  &&& iSPLS-Hetero$_S$ & 10.113 (2.818) & 0.988 (0.062) & 0.895 (0.011)& \\

		\hline
			
		\end{tabular}
	}
	}
\end{center}
\end{table}

\begin{table}
	\caption{Simulation results for Scenario 4($M=4, p=100$)}
	\begin{center}
	\label{Table4}
	\resizebox{125mm}{50mm}{
	\centering{\scriptsize
		\begin{tabular}{llllllllllll}
		\hline\
&$\rho$&$n_l$&Method&MSPE&Sensitivity&Specificity& \\
		       \hline
		
		       &0.2&40& meta-PLS & 203.530 (25.691) & 1 (0) & 0 (0)&\\
		                  &&& meta-SPLS & 92.530 (21.452) & 0.580 (0.105) & 0.880 (0.096)&\\
		                  &&& pooled-SPLS & 174.432 (14.915) & 0.491 (0.300) & 0.608 (0.288)&\\
		                  &&& iSPLS-Homo$_M$ & 100.245 (14.246) & 0.852 (0.067) & 0.404 (0.084)&\\
		                  &&& iSPLS-Homo$_S$ & 98.013 (14.990) & 0.851 (0.064) & 0.403 (0.073)&\\
		                  &&& iSPLS-Hetero$_M$ & 79.508 (19.775) & 0.633 (0.111) & 0.918 (0.026)&\\
		                  &&& iSPLS-Hetero$_S$ & 81.041 (19.177) & 0.626 (0.106) & 0.920 (0.025)&\\

		                 \hline
                        & 0.2&120& meta-PLS & 233.403 (41.459) & 1 (0) & 0 (0)&\\
		                  &&& meta-SPLS & 26.850 (5.595) & 0.689 (0.060) & 0.994 (0.067)&\\
		                  &&& pooled-SPLS & 155.342 (12.988) & 0.500 (0.098) & 0.717 (0.041)&\\
		                  &&& iSPLS-Homo$_M$ & 42.005 (5.538) & 0.914 (0.047) & 0.496 (0.055)&\\
		                  &&& iSPLS-Homo$_S$ & 41.962 (5.566) & 0.913 (0.047) & 0.498 (0.063)&\\
		                  &&& iSPLS-Hetero$_M$ & 24.120 (4.955) & 0.878 (0.076) & 0.925 (0.025)&\\
		                  &&& iSPLS-Hetero$_S$ & 24.177 (4.865) & 0.88 (0.077) & 0.926 (0.018)&\\
		          \hline
		       &0.7&40& meta-PLS & 542.745 (91.125) & 1 (0) & 0 (0)&\\
		                  &&& meta-SPLS & 69.596 (22.876) & 0.654 (0.089) & 0.974 (0.024)&\\
		                  &&& pooled-SPLS & 357.967 (34.464) & 0.401 (0.236 ) & 0.753 (0.219)&\\
		                  &&& iSPLS-Homo$_M$ & 100.322 (17.976) & 0.937 (0.055) & 0.437 (0.067)&\\
		                  &&& iSPLS-Homo$_S$ & 97.774 (20.713) & 0.937 (0.054) & 0.436 (0.069)&\\
		                  &&& iSPLS-Hetero$_M$ & 66.089 (16.784) & 0.904 (0.083) & 0.776 (0.041)&\\
		                  &&& iSPLS-Hetero$_S$ & 64.131 (16.378) & 0.904 (0.089) & 0.771 (0.024)&\\		                 
		                  \hline
		      &0.7&120& meta-PLS & 636.34 (73.501) & 1 (0) & 0 (0)&\\
		                  &&& meta-SPLS & 35.067 (8.960) & 0.872 (0.015) & 0.954 (0.057)&\\
		                  &&& pooled-SPLS & 331.250 (9.337) & 0.469 (0.110) & 0.773 (0.075)&\\
		                  &&& iSPLS-Homo$_M$ & 56.381 (11.017) & 0.992 (0.047) & 0.465 (0.018)&\\
		                  &&& iSPLS-Homo$_S$ & 56.234 (11.021) & 0.993 (0.047) & 0.461 (0.019)&\\
		                  &&& iSPLS-Hetero$_M$ & 31.622 (8.855) & 0.943 (0.066) & 0.913 (0.016)&\\
		                  &&& iSPLS-Hetero$_S$ & 30.625 (8.501) & 0.943 (0.063) & 0.911 (0.017)&\\	     
		                  	          
		\hline
			
		\end{tabular}
	}
	}
\end{center}
\end{table}

To sum up, under the homogeneity cases, iSPLS-Homo$_M$ and iSPLS-Homo$_S$ have the most favourable performance, and under the heterogeneity cases, iSPLS-Hetero$_S$ and  iSPLS-Hetero$_M$ outperform the others. It is also interesting to observe that the performance of the constructed penalties depends on the degree of similarity across datasets. For example, in Table 2, iSPLS-Homo$_S$ (iSPLS-Hetero$_S$), with a less stringent penalty, has relatively lower MSPEs than iSPLS-Homo$_M$ (iSPLS-Hetero$_M$), while in Table 1, it is the other way around. This comparison suggests the sensibility of the proposed contrasted penalization.

\section{Data analysis}
\subsection{Analysis of cutaneous melanoma data}
We analyze three datasets from the TCGA cutaneous melanoma (SKCM) study, corresponding to different tumor stages, with 70 samples in stage 1, 60 in stage 2, and 110 in stage 3 and 4.  Studies have been conducted on Breslow thickness, an important prognostic marker, which is regulated by gene expressions. However, most of these studies use all samples from different stages together. Exploratory analysis suggests that beyond similarity, there also exists considerable variation across the three stages. The number of gene expression measurements contained in these three datasets is 18947 in all. To generate more accurate results with quite limited samples, we conduct our analysis based on the result of \citet{Sun2018community}, in which they develop a Community Fusion (CoFu) approach to conduct variable selection while taking account into the network community structure of omics measurements. After undergoing procedures including the unique identification of genes, matching of gene names with those in the SKCM dataset, a supervised screening, network construction and community identification, a total of 21 communities, with 126 genes, are identified as associated with the response using their proposed CoFu method, and are used here for downstream analysis.

We apply the proposed integrative analysis methods and their competitors, meta-analysis, and pooled analysis. It is found that the identified variables vary across methods and stages. For example, Figure 1 shows the estimation results for genes in community 3, 5 and 42, in which each row corresponds to one dataset.  We can see that, for one specific set, although results generated by different methods vary to each other, they share some nonzero loadings in common, and the number of overlapping genes identified by different methods are summarized in Table 5. Genes identified by iSPLS-Hetero shown in Figure 1 demonstrate the stage-specific feature in a specific community, thereby indicating the difference across tumor stages. 

\begin{figure*}[!t]
\centering
\includegraphics[ height=3.0in, width=4.80in]{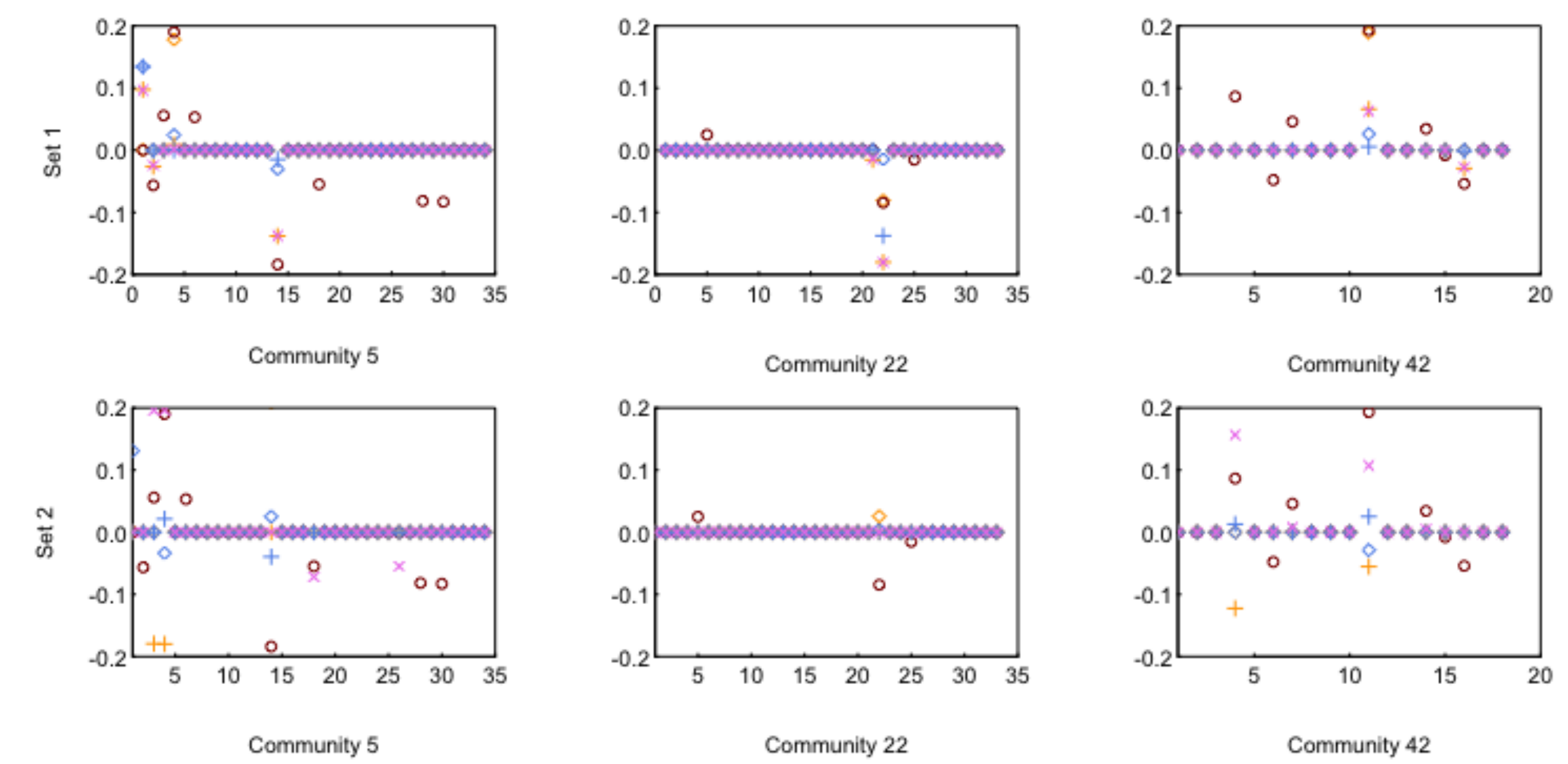}  
\caption{Analysis of the TCGA SKCM data.  Rhombus and cross in blue and orange correspond to iSPLS-Homo and Hetero with magnitude and sign penalties, respectively. Pink cross and red circle correspond to meta-SPLS and pooled-SPLS.}
\label{fig1}
\end{figure*}
  
To evaluate prediction performance and stability of identification, we first randomly split each dataset into 75\% for training and 25\% for testing.  Then, estimation results are generated by the training set and used to make a prediction for the testing set. The root mean squared error (RMSE) is used to measure prediction performance. Furthermore, for each gene, we compute its observed occurrence index (OOI) \citep{huang2010variable}, that is, its probability of being identified in 100 resamplings. The results of RMSEs and OOIs for each method are shown in Table 6, which suggests the stability of our proposed methods as well as their competitive performance compared to the alternatives.

\subsection{Analysis of lung cancer data}
We collect two lung cancer datasets, on Lung Adenocarcinoma (LUAD) and Lung Squamous Cell Carcinoma (LUSC), with sample sizes equal to 142 and 89, respectively. Studies have been conducted to analyze FEV1, which is a measure of lung function, and its relationship with gene expressions, using two datasets, however, separately. Since both Adenocarcinoma and Squamous Cell Carcinoma are non-small cell lung carcinomas, we may expect a certain degree of similarity between them. With the consideration on both difference and similarity, we apply our proposed integrative methods on these two datasets. Our analysis focuses on 474 genes in 26 communities, which are identified as associated with the response, based on the results of  \citet{Sun2018community}. 

We perform the same procedure as described above. The identified variables vary across methods and datasets. To better illustrate the estimation results, Figure 2 shows the behaviors of three communities identified by the above methods, from which we can easily see both the similarities and differences between these two datasets. Stability and prediction performance evaluation are conducted by computing the RMSEs and OOIs from 100 resamplings, following the same procedure as described above. Overall results are summarized in Table 5-6, and the iSPLS methods have relatively lower RMSEs and higher OOIs than the other methods. 

\begin{figure*}[!t]
\centering
\includegraphics[ height=3.60in, width=4.8in]{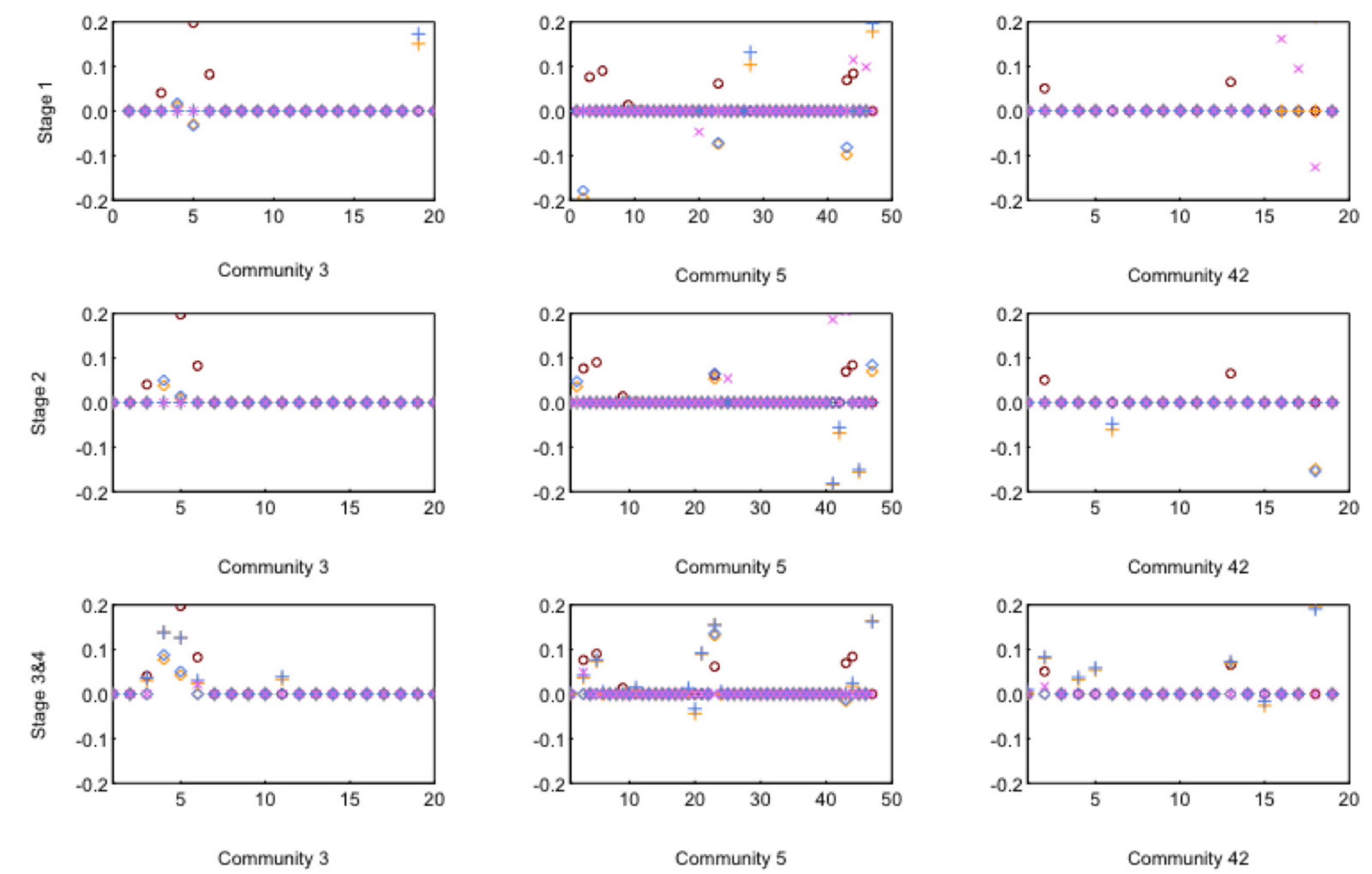}  
\caption{Analysis of the TCGA lung cancer data.  Rhombus and cross in blue and orange correspond to iSPLS-Homo and Hetero with magnitude and sign penalties, respectively. Pink cross and red circle correspond to meta-SPLS and pooled-SPLS.}
\label{fig 2}
\end{figure*} 

\begin{table*}
\centering{
	\caption{Data analysis: numbers of overlapping genes identified by different methods.}
	\label{Table 5}
	\resizebox{120mm}{30mm}{
		\begin{tabular}{lllllll}
		\hline

&&&&iSPLS&&\\
\cline{4-7}
Method&pooled-SPLS &  meta-SPLS &Homo$_M$ & Homo$_S$ &Hetero$_M$&Hetero$_S$ \\
       \hline	
SKCM data &&&&& &\\	
		pooled-SPLS& 100 & 34&20  &21   &51 &53    \\
		   meta-SPLS & &107   & 28 & 29 & 71&72 \\
		    iSPLS-Homo$_M$ &   & & 45 &45   &45  &45 \\
		    iSPLS-Homo$_S$ &   &   &   & 46 &46 &46 \\
		    iSPLS-Hetero$_M$ &   &   &   &&83  &75  \\
		     iSPLS-Hetero$_S$ &   &   &   &  & & 89\\
		       \hline

Lung cancer data	&&&&&&\\
		      pooled-SPLS &145 &78 &37  &40 &66 &51    \\
		     meta-SPLS & &  92 &39  & 42 &76 &58 \\
		    iSPLS-Homo$_M$ &   &   & 39& 39 & 38&35 \\
		    iSPLS-Homo$_S$ &   &  & &   42& 40 &36  \\
		    iSPLS-Hetero$_M$ &   &   &   &  &66 &58 \\
		     iSPLS-Hetero$_S$ &   &   &   &  & &72 \\
	                 		               		\hline
		\end{tabular}
		}
		}
\end{table*}

\begin{table*}
\centering{
	\caption{Data analysis: RMSEs and the median of OOIs of different methods}
	\label{Table 6}
	\resizebox{120mm}{15mm}{
		\begin{tabular}{lllllll}
		\hline\
&&&&iSPLS& & \\
\cline{4-7}
&pooled-SPLS &meta-SPLS &Homo$_M$ & Homo$_S$ &Hetero$_M$& Hetero$_S$ \\
	\hline
	SKCM data	&&&&& & \\		
	RMSE & 6.210 &4.046&4.202 & 4.163 &3.202&  3.135 \\
	OOI(Median) & 0.76 & 0.75 & 0.80 &0.80 &0.77&0.78\\
	 \hline
	Lung cancer data	  &&&&& & \\
	RMSE & 32.367  &27.837 & 22.269 &20.412  & 21.019&   20.318 \\
	OOI(Median) & 0.71 & 0.73 & 0.78 &0.78 &0.76&0.75\\
	 \hline			
		\end{tabular}
		}
		}
\end{table*}

\section{Discussion}
PLS regression has been promoted in ill-conditioned linear regression problems that arise in several disciplines such as chemistry, economics, medicine, and psychology. In this study, we propose an integrative SPLS (iSPLS) method, which conducts the integrative analysis of multiple independent datasets based on the SPLS technique. This study significantly extends the novel integrative analysis paradigm by conducting a dimension reduction analysis. An important contribution is that, to promote similarity across datasets more effectively, two contrasted penalties have been developed. Under both the homogeneity and heterogeneity models, we develop the magnitude-based contrasted penalization and sign-based contrasted penalization. We develop effective computational algorithms for the proposed integrative analysis. For a variety of model settings, simulations demonstrate satisfactory performance of the proposed iSPLS method. The application to TCGA data suggests that magnitude-based iSPLS and sign-based iSPLS do not dominate each other, and are both needed in practice. The stability and prediction evaluation provides some support to the validity of the proposed method.

This study can be potentially extended in multiple directions. Apart from PLS, integrative analysis can be developed based on other dimension reduction techniques, such as CCA, ICA, and so on. For selection, the MCP penalty is adopted and can be potentially replaced with other two-level selection penalties. Integrative analysis can be developed based on SPLS-SVD. Moreover, iSPLS is applicable to non-linear frameworks such as generalized linear models and survival models. In data analysis, both the magnitude-based iSPLS and sign-based iSPLS have applications far beyond this study. 


%
%

\bibliographystyle{apa}
\bibliography{iSPLS}

\newpage
\appendix

\subsection*{Algorithms}

\subsubsection*{iSPLS with 2-norm group MCP with Magnitude-based contrasted penalty}  
We adopt a similar computational algorithm as Algorithm 1. The key difference lies in Step 2(b), solving $c^{(l)}$ with fixed $w_{[t]}^{(l)}$. Consider the homogeneity model with magnitude-based contrasted penalty (iSPLS-Homo$_M$), we have the following problem 
  \begin{equation}
  \label{eq:5.3}
  \begin{aligned}
\min \limits_{c^{(l)}} & \sum_{l=1}^{L} \frac{1}{2n_l^2} \left( \left\|Z^{(l)\top}c^{(l)}-Z^{(l)\top}w_{[t]}^{(l)}\right\|_{2}^{2}
+\lambda \left \|c_1^{(l)}\right \|_2^2 \right) 
+\sum_{j=1}^{p} \rho \left(\left \|{c_{j}}\right \|_2; \mu_1, a\right)\\
&+\frac{\mu_2}{2} \sum_{j=1}^p\sum_{l^\prime \neq l}\left(c_{j}^{(l)}-c_{j}^{(l^\prime)}\right)^2.
  \end{aligned}
\end{equation}
     
For $j=1,\dots,p$, given the group parameter vectors $c^{(l)}_{k} $$(k \not = j) $ fixed at their current estimates $c^{(l)}_{k,[t-1]}$, minimize the objective function (\ref{eq:5.3}) with respect to $c^{(l)}_{j}$. After conducting the same procedures as those in Section 2.3.1, this problem is equivalent to minimizing
  \begin{equation}
  \label{eq:5.4}
  \begin{aligned}
 \frac{1}{2} c^{(l)2}_{j}-w_{[t]}^{(l)\top}Z^{(l)}Z^{(l)\top}_{j}c^{(l)}_{j} +\dot \rho \left(\left\|c_{j,[t-1]}\right\|_2; \mu_{1}, a\right)\left\| c_{j}\right\|_2
 +\frac{\mu_2^*}{2} \sum_{l^\prime \neq l}\left(c_{j}^{(l)}-c_{j}^{(l^\prime)}\right)^2.
 \end{aligned}
  \end{equation}
It can be shown that the minimizer of (\ref{eq:5.4}) is
 \begin{equation*}
  c^{(l)}_{j,[t]}=\frac{\left(\left\|S_{j}\right\|_2-\dot \rho(\left\|c_{j,[t-1]}\right\|_2; \mu_{1}, a)\right)_{+}S^{(l)}_{j}}{(1+\mu_2^*(L-1))\left\|S_{j}\right\|_2},
  \end{equation*}
  where
   $ S_{j}^{(l)}=\sum_{m=1}^{p}\sum_{i=1}^{q}w^{(l)}_{m}Z^{(l)}_{mi}Z^{(l)}_{ji}+\mu_2^*\sum_{l^{\prime} \neq l}c_{j,[t-1]}^{(l^\prime)},$ and $\left\|S_{j}\right\|_2 =\sqrt {\sum^{L}_{l=1} S^{(l)2}_j}$.

\subsubsection*{iSPLS with composite MCP with Magnitude-based contrasted penalty} 
Under the heterogeneity model with Magnitude-based contrasted penalty (iSPLS-Hetero$_M$), 
  \begin{equation}
  \label{eq:5.5}
  \begin{aligned}
\min \limits_{c^{(l)}} &\sum_{l=1}^{L} \frac{1}{2n_l^2} \left( \left\|Z^{(l)\top}c^{(l)}-Z^{(l)\top}w_{[t]}^{(l)}\right\|_{2}^{2}
+\lambda\left \|c^{(l)}\right \|_2^2\right) 
+\sum_{j=1}^p \rho \left( \sum_{l=1}^L \rho\left(| c_{j}^{(l)} |;\mu_1, a\right); 1,b \right)\\
&+\frac{\mu_2}{2} \sum_{j=1}^p\sum_{ l^\prime \neq l }\left(c_{j}^{(l)}-c_{j}^{(l^\prime)}\right)^2.
  \end{aligned}
 \end{equation} 
  
Take the first order Taylor expansion approximation about $c_{j}^{(l)}$ for the first penalty, with $c_{k}^{(l)}(k\not=j)$ fixed at their current estimates $c^{(l)}_{k,[t-1]}$, and conduct the same procedure to the second penalty as in Section 2.3.1. Then the objective function (\ref{eq:5.5}) is approximately equivalent to minimizing  
\begin{equation}
  \frac{1}{2} c_{j}^{(l)2}-w_{[t]}^{(l)\top}Z^{(l)}Z^{(l)\top}_{j}c^{(l)}_{j} +\alpha_{jl} | c^{(l)}_{j} | +\frac{\mu_2^*}{2} \sum_{l^\prime \neq l }\left(c_{j}^{(l)}-c_{j}^{(l^\prime)}\right)^2,~
\end{equation}
where 
$\alpha _{jl}=\dot \rho \left(\sum_{l=1}^L\rho( | c_{j,[t-1]}^{(l)}|; \mu_1, a); 1, b\right)\dot\rho \left( | c^{(l)}_{j,[t-1]} |; \mu_{1}, a\right)$.

Thus, $c^{(l)}_{j,[t]}$ can be updated as follows: For $l=1,\dots,L$,
\begin{itemize}
\item[1.] Initialize $r=0$ and $  c^{(l)}_{j,[r]}= c^{(l)}_{j,[t-1]}$.
\item[2.] Update $r=r+1$. Compute:
 \begin{equation*}
  c^{(l)}_{j,[r]}=\frac{\text{sgn}\left(S_{j,[r-1]}^{(l)}\right)\left(|S_{j,[r-1]}^{(l)}|-\alpha_{jl}\right)_{+}}{(1+\mu_2^*(L-1))},
 \end{equation*}
  where $$ S_{j,[r-1]}^{(l)}=\sum_{m=1}^{p}\sum_{i=1}^{q}w^{(l)}_{m}Z^{(l)}_{mi}Z^{(l)}_{ji}+\mu_2^*\sum_{l^{\prime}\not=l}c_{j,[t-1]}^{(l^{\prime})},$$ and
   $\alpha _{jl}=\dot \rho \left(\sum_{l=1}^{L}\rho( | c_{j,[r-1]}^{(l)}|; \mu_1, a); 1, b\right)\dot\rho \left( | c^{(l)}_{j,[r-1]} |; \mu_{1}, a\right).$
\item[3.] Repeat Step 2 until convergence. The estimate at convergence is $c^{(l)}_{j,[t]}$.
\end{itemize}

\subsubsection*{iSPLS with 2-norm group MCP with the sign-based contrasted penalty}
Consider the homogeneity model with sign-based contrasted penalty (iSPLS-Homo$_S$). 
  \begin{equation}
  \label{eq:5.1}
\begin{aligned}
  \min \limits_{c^{(l)}} &\sum_{l=1}^{L} \frac{1}{2n_l^2} \left( \left\|Z^{(l)\top}c^{(l)}-Z^{(l)\top}w_{[t]}^{(l)}\right\|_{2}^{2}
  +\lambda \left \|c^{(l)}\right \|^2_2 \right)
  +\sum_{j=1}^{p} \rho\left(\left \|{c_{j}}\right \|_2; \mu_1, a\right)\\
&+\frac{\mu_2}{2} \sum_{j=1}^p\sum_{l^\prime \neq l}\left\{\text{sgn}(c_{j}^{(l)})-\text{sgn}(c_{j}^{(l^\prime)})\right\}^2.
\end{aligned}
\end{equation}  
  
For $j=1,\dots,p$, following the same procedure in Section 2.3.1, we have the following minimization problem 
 \begin{equation}
   \label{eq:5.2}
  \begin{aligned}
   \frac{1}{2}& c^{(l)2}_{j}-w_{[t]}^{(l)\top}ZZ^{(l)\top}_{j}c^{(l)}_{j} +\dot \rho \left(\left\|c_{j,[t-1]}\right\|_2; \mu_{1}, a \right)\left\| c_{j}\right\|_2 \\
&+ \frac{\mu_2^*}{2}\sum_{ l^\prime \neq l} \Bigg( \frac{c_{j}^{(l)}}{\sqrt{c_{j}^{(l)2}+\tau^2}}-\frac{c_{j}^{(l^\prime)}}{\sqrt{c_{j}^{(l^\prime)2}+\tau^2}} \Bigg)^2.
  \end{aligned}
\end{equation}

It can be shown that the minimizer of (\ref{eq:5.2}) is
  \begin{equation*}
  c^{(l)}_{j,[t]}=\frac{\left(\left\|S_{j}\right\|_2-\dot \rho(\left\|c_{j,[t-1]}\right\|_2; \mu_{1}, a)\right)_{+}S^{(l)}_{j}}{(1+\mu_2^*(L-1))/( c^{(l)2}_{j,[t-1]}+\tau^{2} )\left\|S_{j}\right\|_2},
 \end{equation*}
  where
  \begin{equation}
  \begin{aligned}
   S_{j}^{(l)}&=\sum_{m=1}^{p}\sum_{i=1}^{q}w^{(l)}_{m}Z^{(l)}_{mi}Z^{(l)}_{ji}+\frac{\mu_2^*}{\sqrt{c^{(l)2}_{j,[t-1]}+\tau^{2}}}\sum_{l^{\prime}\not=l}\frac{c_{j,[t-1]}^{(l^{\prime})}}{\sqrt{c_{j,[t-1]}^{(l^{\prime})2}+\tau^{2}}},
  \end{aligned}
  \end{equation}  
  and $\left\|S_{j}\right\|_2 =\sqrt {\sum^{L}_{l=1} (S^{(l)}_{j})^2}$.

 Thus, $c^{(l)}_{j,[t]}$ can be updated as follows: For $l=1,\dots,L$
 \begin{itemize}
\item[1.] Initialize $r=0$ and $  c^{(l)}_{j,[r]}= c^{(l)}_{j,[t-1]}$
\item[2.] Update $r=r+1$. Compute:
\begin{equation*}
  c^{(l)}_{j,[r]}=\frac{\left(\left\|S_{j,[r-1]}\right\|_2-\dot \rho\left(\left\|c_{j,[r-1]}\right\|_2; \mu_{1}, a\right)\right)_{+}S^{(l)}_{j,[r-1]}}{(1+\mu_2^*(L-1))/( c^{(l)2}_{j,[r-1]}+\tau^{2} )\left\|S_{j,[r-1]}\right\|_2}.
   \end{equation*}
 \item[3.] Repeat Step 2 until convergence. The estimate at convergence is $c^{(l)}_{j,[t]}$.
\end{itemize}
\clearpage{}

\end{document}